# Global Distribution of Serpentine on Mars

**A. Emran**[1*], **J. D. Tarnas**[2], and **K. M. Stack**[1]

[1] NASA Jet Propulsion Laboratory, California Institute of Technology, CA 91109, USA
[2] Blue Origin Space Resources Program, Los Angeles, CA, USA

[*] Corresponding author: A. Emran (al.emran@jpl.nasa.gov)

**Abstract**

The distribution and origin of serpentine on Mars can provide insights into the planet's aqueous history, habitability, and past climate. In this study, we used dynamic aperture factor analysis/target transformation applied to 15,760 images from the Compact Reconnaissance Imaging Spectrometer for Mars, followed by validation with the radiance ratio method, to construct a map of Mg-serpentine deposits on Mars. Although relatively rare, Mg-serpentine was detected in diverse geomorphic settings across Noachian and Hesperian-aged terrains in the southern highlands of Mars, implying that serpentinization was active on early Mars and that multiple formation mechanisms may be needed to explain its spatial distribution. We also calculated the amount of $H_2$ produced during the formation of the observed deposits and conclude that serpentinization was likely more widespread on Mars than indicated by the observed distribution.

**Plain Language Summary**

Determining the location and origin of serpentine minerals on Mars can tell us about the past climate and history of water on Mars, as well as its potential to support ancient life. In this study, we used data mining methods to search for Mg-rich serpentine in 15,760 images from the Compact Reconnaissance Imaging Spectrometer for Mars. Serpentine was detected in a variety of settings within the oldest Martian rocks, implying that serpentinization was active on early Mars. Although Mg-rich serpentine appears rare on Mars, the formation of serpentine minerals was likely widespread.

*Key points:*

- Mg-rich serpentine was identified in 43 CRISM images, covering a surface area of ~1.46 km$^2$ on Mars.
- Serpentine occurs in Noachian- and Hesperian-aged terrains in a variety of geomorphic settings.
- Despite the observed paucity of serpentine on Mars, its formation was likely widespread.

**1. Introduction**

Serpentinization is the process by which serpentine is produced via the oxidation of olivine and pyroxene by liquid water (e.g., McCollom, 2016):

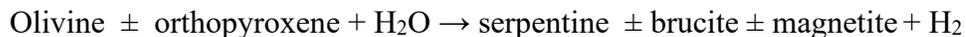

Olivine ± orthopyroxene + H$_2$O → serpentine ± brucite ± magnetite + H$_2$

Serpentinization on Earth occurs in settings such as hydrothermal systems and tectonically active margins on the ocean floor (e.g., Schulte et al. 2006; Amador et al. 2017). Chemolithotrophic microbial communities near such sites rely on H$_2$ and CH$_4$ nutrient sources (Schrenk et al. 2013; McCollom and Donaldson, 2016). On Mars, where serpentinization is favored thermodynamically (e.g., Oze and Sharma, 2005; McCollom et al. 2022), serpentinized rocks are a compelling astrobiology target for *in situ* Martian exploration (e.g., Vance and Melwani Daswani, 2020; Steele et al. 2022). Abiotic H$_2$ and CH$_4$ produced via serpentinization may also have contributed significantly to the warming of the planet (e.g., Wordsworth et al. 2021). Understanding the extent and timing of serpentinization has important implications for the astrobiological potential and paleoclimate of Mars. Moreover, serpentinites on Earth are often metal-rich, containing Ni- and Cr-rich ores, and could represent sites for future metal extraction.

Serpentine has been reported on Mars using data from the Compact Reconnaissance Imaging Spectrometer for Mars (CRISM) at local to global scales (e.g., Ehlmann et al. 2009, 2010; Bultel et al. 2015; Viviano-Beck et al. 2017; Amador et al. 2017, 2018; Lin et al. 2021). A confident detection of Mg-serpentine requires the presence of characteristic absorption bands at ~1.4, 2.12, 2.32, and 2.52 μm in CRISM data (Ehlmann et al. 2010). However, an artifact that mimics the diagnostic absorption feature for serpentine at 2.12 μm was identified in CRISM I/F data (Leask et al. 2018). A validation method, referred to as the radiance ratio method, was developed to confirm the presence of this 2.12-μm absorption in CRISM radiance data (Leask et al. 2018). To-date, serpentine has been validated using this method within only six CRISM images (Leask et al. 2018; Lin et al. 2021; Tarnas et al. 2021; refer to S1.1).

In this study, we applied dynamic aperture factor analysis and target transformation (DAFA/TT; Lin et al. 2021) to identify and localize candidate serpentine detections within CRISM I/F data. Potential detections were then validated using the radiance ratio method to verify the presence of a 2.12-µm band. We generated an updated global map of detectable Mg-rich serpentine on Mars, investigated the geologic context, terrain age, and nearby mineral assemblages of these detections, and calculated the amount of $H_2$ that could have been generated via serpentinization of these deposits. This analysis seeks to constrain the extent, timing, and processes responsible for serpentinization on Mars.

## 2. Observations and Methods

To localize serpentine spectra, we examined a total of 15,760 CRISM Targeted Reduced Data Record images (TRDR, Murchie et al. 2007), including 7885 Full Resolution Targeted (FRT; ~18 m/pixel), 3738 Full Resolution Short (FRS; ~18 m/pixel), 2785 Half Resolution Long (HRL; ~36 m/pixel), and 1352 Half Resolution Short (HRS; ~36 m/pixel) observations. These CRISM images range in latitude from 90º N to 90º S and cover ~1.2% of the Martian surface. We applied a standard CRISM processing pipeline for atmospheric and photometric calibrations to all image data including a correction for atmospheric gas using the volcanic-scan correction routine (McGuire et al. 2009).

The DAFA/TT method applies factor analysis/target transformation (Malinowski, 1991)—the extraction of spectral endmembers that impart variance onto a set of spectra, then combination of those endmembers through unconstrained linear least squares fitting to achieve the lowest root mean square error to a library spectrum of a specific mineral—in windows of pixels that move across a hyperspectral image (Lin et al. 2021; refer to S1.2). DAFA/TT was applied to all calibrated CRISM images between the wavelength range of 1-2.6 µm (Lin et al. 2021; Tarnas et al. 2021) to identify candidate serpentine pixel clusters. For each pixel within a cluster, we calculated two I/F spectral ratios using the same numerator spectrum divided by, respectively: (1) a "bland" denominator spectrum calculated as a simple median from the same detector column (Leask et al. 2018), and (2) the spectrum of nearby "bland" pixels, selected using the method of Plebani et al.

(2022). The radiance ratio spectra for each pixel in a cluster were extracted following the same approach. Average I/F and radiance ratio spectra were then calculated for the candidate clusters.

To avoid the 2.12-µm artifact in the I/F data, we visually confirmed this absorption band in the radiance ratio spectra of candidate serpentine detections. The robustness of each serpentine detection was evaluated on a scale of 1 (low/weak) to 5 (high/best) based on the position, strength, and shape of the absorption bands at ~ 1.38, 2.12, 2.32, and 2.52-µm in both I/F and radiance spectra (e.g., Fig. S1). We investigated the geologic and geomorphic context of each serpentine detection using CTX and THEMIS mosaics (Dickson et al. 2023; Edwards et al. 2011) and evaluated nearby mineral assemblages (see S1.3). To explore the effect of dust, we compared the spatial distribution of validated serpentine detections with Ruff and Christensen's (2002) TES dust cover index map.

The age of validated serpentine deposits was estimated using the geologic map of Tanaka et al. (2014), recognizing that this map provides surface exposure age estimates of the units in which serpentine is detected, not necessarily the formation age of serpentine itself. We also calculated ranges for the amount of molecular $H_2$ (mol) generated during the formation of the validated deposits, assuming a serpentinization depth up to 10 km, and assuming a range of $H_2$ production rates from an olivine-rich rock (Tutolo and Tosca, 2023) analogous to crust on Mars (see S1.4). We consider 10 km to be the maximum possible depth of serpentinization since the water ceases to flow at that depth (e.g., Oze and Sharma, 2005). Moreover, a hydrothermal fluid-rich region has been estimated to exist >5km deep in the subsurface, with hydrated minerals detected that exhumed from depths exceeding 6km (e.g., Michalski et al. 2013).

## 3. Results

We identified serpentine in 43 CRISM images across the surface of Mars, including 32 images containing new, previously unrecognized detections (Fig. 1, Table S1, Fig. S2). Fifteen detections were made with high confidence (confidence levels 4 and 5), 18 with medium confidence (confidence level 3), and 10 with low confidence (confidence levels 1 and 2) (Table S1). Serpentine detections are constrained latitudinally between 40º N to 40º S, and no serpentine was found between longitudes 90º E to 180º E (Fig. 1), which corresponds to a dustier region of Mars (Fig. 3). Many serpentine detections occur in the less dusty regions between 60-90°E longitude (Fig. 1 and 3d). However, the least dusty longitudes on Mars (~60°W-0°E) show fewer detections. Serpentine detections occur at Nili Fossae and Syrtis Major, Tyrrhena Terra north of Hellas Planitia, Mawrth Vallis including Margaritifer Terra, Valles Marineris, Thaumasia Highlands, Terra Sirenum, and mélange terrains at Claritas Rise (Fig. 1, Fig. S2). Of these regions, Nili Fossae and its surrounding areas exhibit the highest number of serpentine detections, 18 of the 43 CRISM images (Fig. S3).

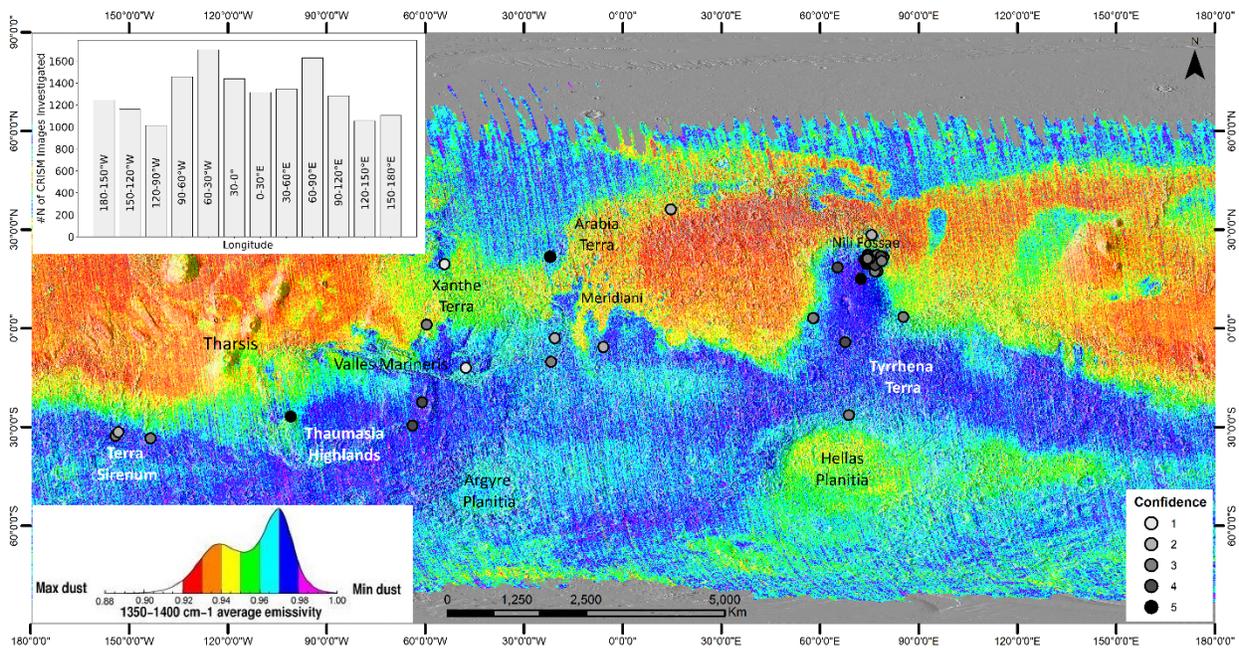

**Fig. 1**: Global distribution of serpentine on Mars overlain on the TES dust cover index color map with the background of MOLA shaded global mosaic (Ruff and Christensen, 2002). Dust cover index values (lower left) represent the average surface emissivity where lower values indicate higher dust. Serpentine detections are coded according to confidence level. Upper left inset shows number of images investigated in 30° longitudinal bin.

Figure 2 shows two new serpentine detections: one on crater ejecta in Terra Sirenum (Fig. 2a) and one on a bedrock cliff in the Thaumasia Highlands (Fig. 2e). CRISM observation HRL0000860C (center coordinate 32.7°S, 205.5°E; Fig. 2a) is in a large basin in Terra Sirenum, where chloride salts were mobilized and deposited by near-surface liquid waters (e.g., Glotch et al. 2010; Davila et al. 2011). The occurrence of serpentine in such a setting is unique and exceptionally relevant to astrobiology because such environments on Earth are known to host life (e.g., Wierzchos et al. 2006). CRISM observation FRS0003F22E (center coordinate 22.5°S, 298.77°E; Fig. 2e) is in the highly deformed Noachian crust surrounding the eastern Thaumasia Highlands at Coprates Rise (Viviano-Beck et al. 2017). Serpentine in this region (Fig. 2f) may indicate formation linked to regional crustal heat flow and fluid circulation (Viviano-Beck et al. 2017) and later exposed due to uplift and slumping. These examples highlight two of the diverse settings in which serpentine occurs on Mars.

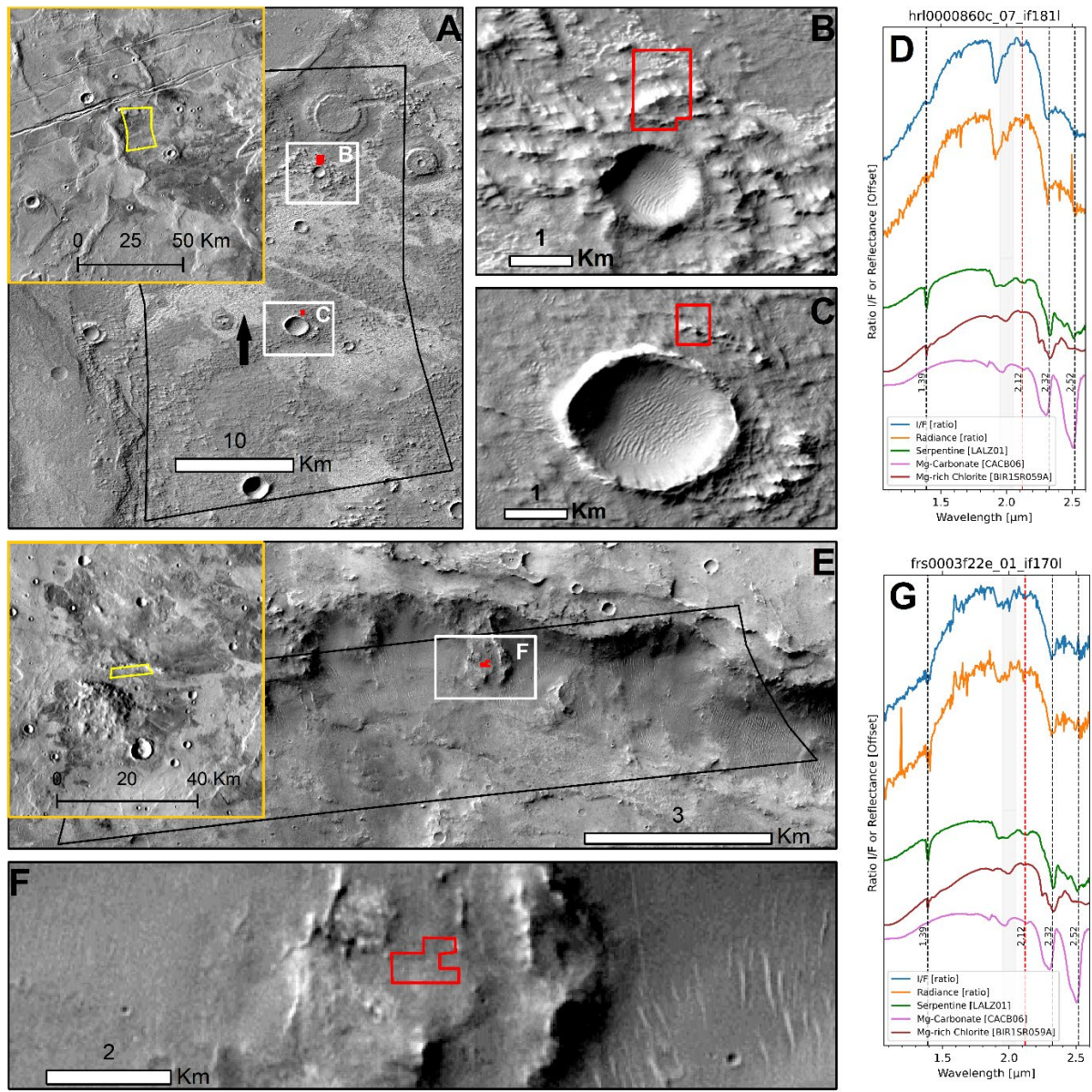

**Fig. 2:** Two new serpentine detections on Mars. a) Black outline represents HRL0000860C footprint overlain on a CTX image. Upper left inset shows the regional context on the THEMIS day IR mosaic. Black arrow points to the light-toned unit corresponding to chloride deposits (Glotch et al. 2010). b) and c) show the local geologic context of two serpentine detections (red polygons) on CTX. d) The average I/F (blue) and radiance (orange) ratio spectra of both pixel clusters showing the presence of a 2.12-µm absorption band plotted with laboratory spectra from the RELAB database (Milliken et al. 2021). e) FRS0003F22E footprint in black overlain on a CTX image. Upper-left inset shows the regional context on the THEMIS day IR mosaic. f) shows the local geologic context of the serpentine detection (red polygons). g) The average I/F (blue plot) and radiance (orange plot) ratio spectra of the detected pixels.

Nineteen serpentine detections occur on "plains," 18 within or around craters, and 6 within valleys (Fig. 3a; Fig. S3). The plains detections exhibit a range of morphologies including rough, relatively smooth, and knobby (chaos) terrains (Fig. 3a). Serpentine was detected on walls and rims, central peaks, ejecta, and crater floors. In valleys, serpentine occurs on both floors and walls. All serpentine deposits identified in this study co-occur with Fe/Mg phyllosilicates within the same CRISM image. Of the 43 images containing serpentine deposits, 36 show evidence of olivine, and 37 show low-Ca pyroxene and Mg- or Ca/Fe-carbonate mineralogy (Fig. 3b). Some serpentine deposits in Terra Sirenum and Nili Fossae occur near chlorites (Table S1).

Of the 43 serpentine detections, 31 occur on Noachian-aged surface units (Tanaka et al. 2014), although nearly 50% of all investigated CRISM images fall within Noachian-aged terrains (Fig. 3c). Among the remaining, we found 2 in Hesperian-Noachian, 4 in Hesperian, 1 in Amazonian-Noachian, and 5 in Amazonian-Hesperian terrain units (Fig. 3c). Serpentine occurring within Amazonian-Hesperian units is found in the fill of Leighton crater, the ejecta of Hargraves crater, and on the floor of the Nili Fossae trough (Table S1). Though around ~8 % of the total CRISM observations fall under Amazonian terrains, none were found via this study's method to contain serpentine.

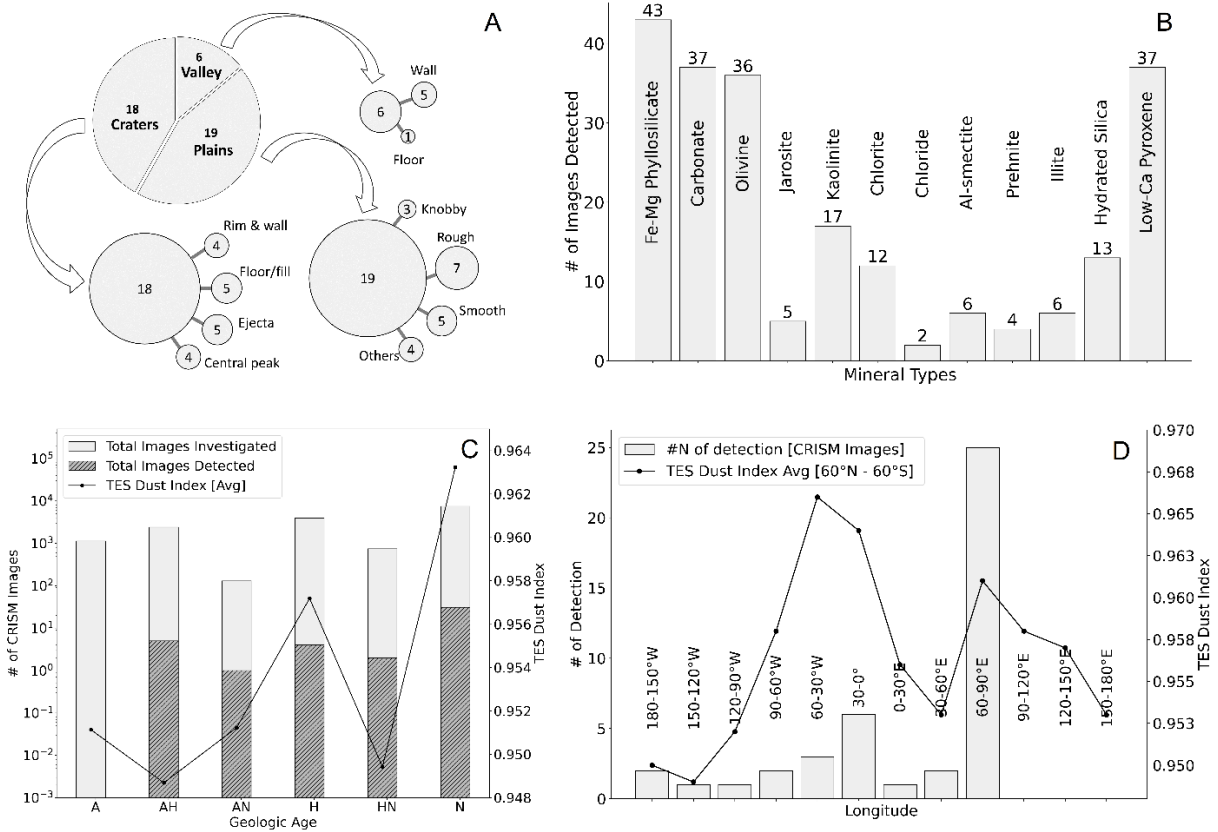

**Fig. 3:** Characteristics of serpentine deposits on Mars. (a) Geologic setting of serpentine deposits. (b) Minerals that co-occur with serpentine deposits. (c) Number of images containing serpentine and all images investigated, binned by the age of units they occur in (Noachian (N), Hesperian-Noachian (HN), Hesperian (H), Amazonian-Noachian (AN), Amazonian-Hesperian (AH), and Amazonian (A)), plotted with the average TES dust index values (right axis). (d) Longitudinal distribution of average TES dust index values and the number of CRISM images containing serpentine. Higher TES dust index values correspond to less dusty regions.

The detected Mg-rich serpentine covers an approximate total surface area of 1.46 km$^2$. If serpentinization extended to a depth of 1 km for this area, the amount of $H_2$ produced during serpentinization would range from 5.11 x 10$^{10}$ mols, assuming 0.01 mol $H_2$/kg production rate (Tutolo and Tosca, 2023), to 4.85 x 10$^{12}$ mols, assuming a 0.95 $H_2$/kg production rate (Tutolo and Tosca, 2023) (Fig. 4). If serpentinization extended to a depth of 10 km (an extreme maximum possibility), the amount of $H_2$ produced during serpentinization would range from 5.11 x 10$^{11}$ mols, assuming 0.01 mol $H_2$/kg production rate (Tutolo and Tosca, 2023), to 4.85 x 10$^{13}$ mols, assuming a 0.95 $H_2$/kg production rate (Tutolo and Tosca, 2023) (Fig. 4).

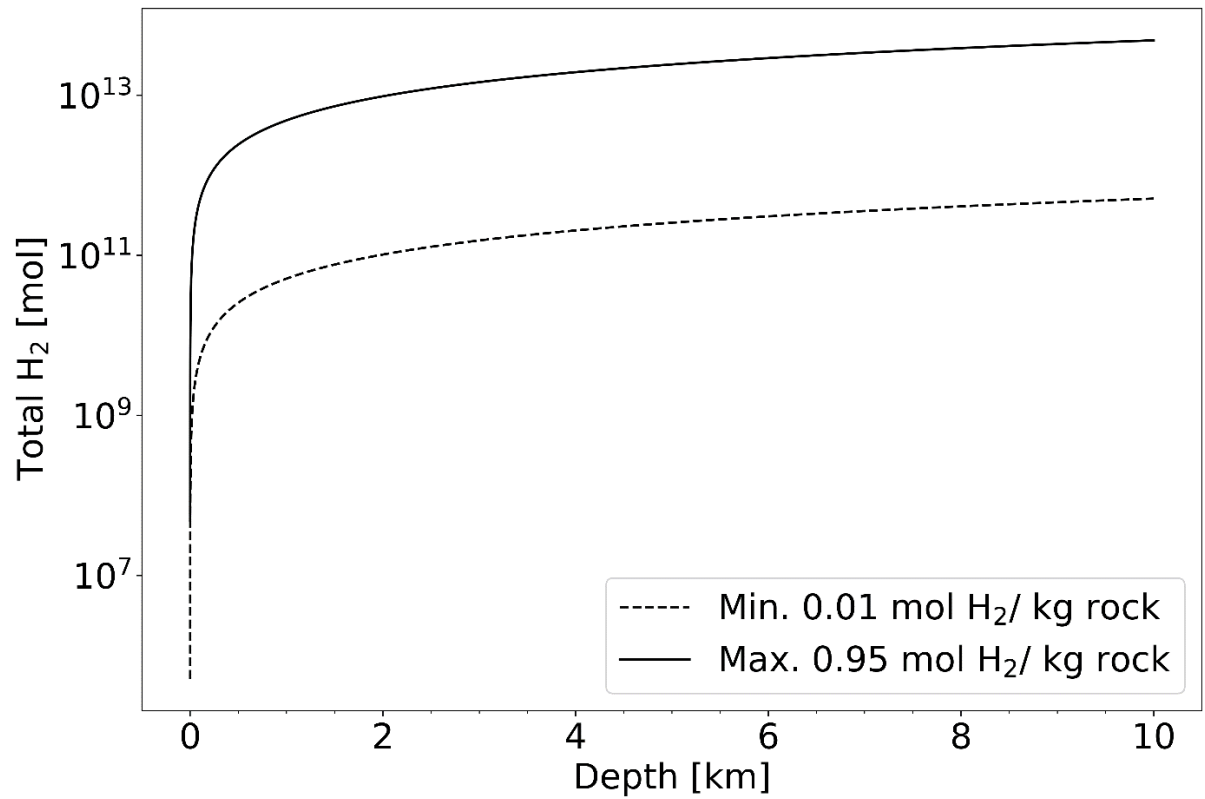

**Fig. 4:** Total amount of molecular $H_2$ (mol) that could have been produced via serpentinization of the validated Mg-serpentine deposits as a function of serpentinization depth. The minimum (min) and maximum (max) production rate (mol/kg of rock) of $H_2$ depends on the degree of serpentinization (Tutolo and Tosca, 2023).

## 4. Discussion

*4.1 Comparison with previous studies*

This study's results generally align with previous studies at the regional scale (e.g., Ehlmann et al. 2010; Amador et al. 2018), although discrepancies between individual CRISM images are common. Using the radiance ratio method, we confirm the presence of serpentine previously identified in six CRISM images in Nili Fossae, Claritas Rise, and Chia and McLaughlin craters (Leask et al. 2018; Lin et al. 2021; Tarnas et al. 2021). A global study by Amador et al. (2018) using the FA/TT method (Malinowski, 1991) found 49 CRISM images containing serpentine. Although 48 of the 49 images identified by Amador et al. (2018) to contain serpentine were candidate outputs from DAFA/TT in this study, we were able to validate only 11 using the radiance ratio technique (Table S1). In Nili Fossae, both Amador et al. (2018) and the DAFA/TT method identified 18 candidate images consistent with serpentine (Fig S2), but only 7 showed the 2.12-µm band in radiance data (Fig. S3). The 4 other detections from Amador et al. (2018) that we validated are in Claritas Rise, north Hellas Planitia, and Leighton and Chia craters (Table S1, Fig. S2).

Although we were unable to validate the presence of serpentine in 38 of Amador et al.'s (2018) images, in several cases, we identified new serpentine detections in images near Amador et al.'s (2018) original detections (Fig. S3). The difference in the detection approach of FA/TT versus DAFA/TT may explain this discrepancy. The FA/TT method used by Amador et al. (2018) identifies individual CRISM images that may contain serpentine but does not localize which pixels contain the serpentine spectra. In contrast, the DAFA/TT method localizes pixels in the image where the serpentine signature originates (Lin et al. 2021), enabling application of the radiance ratio technique to check for the 2.12-µm artifact. For the 38 images of Amador et al. (2018) in which we could not verify serpentine, it is possible that these detections instead identified the known 2.12-µm artifact, or that serpentine is present in the image, but is outside the detection criteria of the DAFA/TT method used here. Nevertheless, with the advantage of localizing mineral detections at the pixel scale, the hyperspectral data mining methods used here could be applied to detect small outcrops of scientifically and economically interesting compounds, such as lunar and asteroidal metals and ices.

*4.2. Age and Formation Mechanisms of Serpentine*

Most serpentine deposits identified in this study occur in Noachian-aged terrains, suggesting that serpentinization occurred predominantly early in Mars's history. However, nearly half of all CRISM observations cover Noachian terrain, which also tends to be less dusty, so there may be an observational bias (Fig. 3). Several serpentine deposits occur in relatively younger Amazonian-Hesperian-aged terrains, but all are in geologic settings related to impact craters (Table S1). This raises the possibility that serpentine originally formed in older Noachian or Hesperian terrains in the subsurface and were later excavated via impact. Eight percent of total CRISM coverage is of Amazonian-aged terrains and we detected no serpentine in these images. Although Amazonian-aged terrains tend to be dustier (Fig. 3), the absence of serpentine detections could indicate that serpentinization did not occur during this period.

The detection of serpentine in a wide variety of geologic settings on Mars suggests that multiple formation mechanisms and distinct hydrothermal environments may be needed to explain the global and local distribution of serpentine. *In situ* formation of serpentine via aqueous alteration of ultramafic rocks is likely at sites like Nili Fossae (e.g., Ehlmann et al. 2010; Viviano et al. 2013; Amador et al. 2017). Elsewhere, the presence of serpentine in crater rims or central peaks raises the possibility of serpentinization in impact-generated hydrothermal systems. For detections in impact ejecta, it is possible that serpentine formed *in situ* at depth and was later excavated during impact. In terrains with little stratigraphic context, like mélange terrains, serpentine could have formed prior, during, or after the deformation of the terrains (Ehlmann et al. 2010), but enhanced fluid flow along fractures and faults in these disrupted terrains could have enabled serpentinization.

In Nili Fossae, the heat source for *in situ* serpentinization may have been related to the nearby Syrtis Major volcanic system (Hiesinger and Head III, 2004), or the formation of the Isidis impact basin. In contrast, serpentine formation at the edges of the Thaumasia Highlands is linked to regional crustal heat flow and fluid circulation, and orogenic processes (Viviano-Beck et al. 2017). Thus, it is unlikely that a single mechanism accounts for formation of all serpentine deposits on Mars. Future study focusing on the local- to regional-scale context and formation of serpentine is warranted.

*4.3. Extent of Serpentinization on Mars*

Mafic and ultramafic terrains are expansive on the Martian surface and serpentinization is thought to be favored thermodynamically on Mars (e.g., Oze and Sharma, 2005; McCollom et al. 2022). However, the Mg-serpentine identified in this study occurs as relatively isolated and small detections totaling just 1.46 km$^2$ within only 43 of 15,760 investigated CRISM images. Based on this observation, we might conclude either that serpentinization was not extensive on Mars or that we are significantly limited in our ability to observe serpentine. Several factors could contribute to the limited detection of serpentine on Mars using orbiter datasets. First, serpentine is a challenging mineral to identify in orbiter spectroscopic data because its diagnostic 2.12-µm band is shallow and close to an atmospheric $CO_2$ absorption band at 2.0-µm (Amador et al. 2018). Sub-pixel mineral mixing could also contribute to the weakening of the shallow 2.12-µm band.

It is also possible that much of the serpentine on Mars is Fe-rich serpentine, which is not distinguishable from other Fe/Mg-phyllosilicates in VIS-NIR data. Fe-rich serpentine minerals are likely to be present on the surface of Mars (Tutolo and Tusca, 2023) given the high iron concentration of the martian basaltic and ultramafic crust compared to terrestrial Mg-rich peridotite (Wade et al. 2017). However, orbiter spectrometers like CRISM cannot be used to identify Fe-rich serpentine given that its characteristic absorption bands overlap with Fe-smectite (nontronite; Calvin and King, 1997) and the 2.12-μm absorption band found in Mg-rich serpentine is absent. This may help explain why the global distribution of detected Mg-serpentine in this study does not align with the distribution of olivine on Mars. For instance, while we detected serpentine in olivine-rich Nili Fossae and Tyrrhena Terra regions, no detection was reported near Argyre Planitia, despite the presence of olivine-rich terrain in the region (Koeppen and Hamilton, 2008; Ody et al. 2013). As Fe-serpentine is indistinguishable from nontronite using orbital data, Tutolo and Tosca (2023) proposed that the extent of Fe-smectite could represent a maximum surface estimate of Fe-serpentine on Mars.

CRISM covers only ~1.2% of the surface area of Mars, raising the possibility that substantial serpentine deposits exist on the surface of Mars that have simply not been observed via the current datasets. While this study almost certainly underestimates the total amount of serpentine present on the surface of Mars, CRISM's targeting bias towards less dusty, ancient regions on Mars where Mg-serpentine could be detected if it were present, suggests that the paucity of serpentine detections is not due solely to limitations of the present orbiter data coverage. However, the possibility remains that there are substantial serpentine deposits present beneath dusty terrains impenetrable by CRISM, that could not be detected even if coverage was more extensive. Moreover, any serpentinization that occurred primarily in the subsurface is observable only when excavated by impact cratering or other tectonic processes.

Serpentinization has been considered an important possible source of $H_2$ and $CH_4$ contributing significantly to warming up early Mars (Chassefière et al. 2013, 2016; Tarnas et al. 2018; McCollom et al. 2022). On Mars, the estimated (averaged over 4.5 Gyr) rate of $H_2$ emission varies from $4.5 \times 10^3$ to $1.48 \times 10^4$ mol s$^{-1}$ (Wordsworth et al. 2021) and the minimum one-year equivalent $H_2$ emission is $1.64 \times 10^4$ mol. To meet this minimum emission level, the serpentinization rate in the subsurface, i.e., the rate at which the reaction front advances at depth, would need to exceed 0.33 km/year if we assume that the 1.46 km$^2$ detected Mg-serpentine had undergone complete alteration at a time with maximum H$_2$ production rate. However, this estimated serpentinization rate is substantially higher than reaction front advance rates of ultramafic rocks on Earth (10$^{-8}$ to 10$^{-4}$ km/year, e.g., Hatakeyama et al. 2017; Beinlich et al. 2020; Leong et al. 2021; Tutolo and Tosca, 2023). Even with an emission rate of $1.64 \times 10^4$ mol year$^{-1}$ H$_2$, the Mg-serpentine deposits observed in this study, if assumed to extend to the maximum depth of ~10 km, could only sustain ~29 years of H$_2$ emission, at odds with the Mars's extended history of aqueous activity (Carr and Head, 2010). Therefore, if serpentinization is indeed an extensive source of H$_2$ emission, serpentine deposits must be far more widespread than what we have detected from orbit.

The apparent paucity of serpentine on Mars could indicate that this mineral once formed extensively early Mars's history but was later altered to clay minerals and carbonates. We found an association between carbonates and serpentine minerals (Fig. 3), providing possible evidence that serpentinization products may have undergone carbonation (Klein and McCollom, 2013; Grozeva et al. 2017; Kelemen and Matter, 2008). The alteration of serpentine into magnesite (Mg-carbonate) and talc via carbonation has also been proposed in Nili Fossae (Viviano et al. 2013; Amador et al. 2017). Furthermore, serpentine has been detected in areas where hydrated minerals are found on the planet (Carter et al. 2013), supporting the possibility that serpentine might have weathered to clay minerals.

## 5. Conclusion

This study provides the most rigorous assessment to-date of the distribution of Mg-serpentine exposed at the Martian surface. The prevalence of serpentine deposits in the southern highlands is evidence of aqueous activity on early Mars. While Mg-serpentine was only rarely detected in less dusty terrains of Noachian-Hesperian age, dust-covered regions of Mars may contain serpentine currently undetectable by orbiter spectrometers. The inability to distinguish Fe-rich serpentine from other Fe/Mg-phyllosilicates with orbiter spectroscopic datasets further limits our ability to constrain and quantify the true extent of serpentinization on Mars. However, our results confirm that serpentinization was an active geochemical process in early Martian history, producing H$_2$ and CH$_4$ that could have contributed to the warming of early Mars. The serpentine deposits identified in this study may have hosted paleoenvironments suitable for microbial communities, making excellent candidates for future *in-situ* Mars rover and lander missions.

**Acknowledgment**

This research was carried out at the Jet Propulsion Laboratory (JPL), California Institute of Technology, under a contract with the National Aeronautics and Space Administration (80NM0018D0004). The work contributing to this paper was not performed by Blue Origin and contains no Blue Origin intellectual property. We acknowledge JPL's High-Performance Computing supercomputer facility, which was funded by JPL's Information and Technology Solutions Directorate. We also acknowledge Annabel Flatland for contributions to the initial processing of data. We acknowledge the CRISM/Mars Reconnaissance Orbiter (MRO) spacecraft missions and PDS Geoscience Node for collecting, storing, and disseminating the data. The authors acknowledge Benjamin Tutolo and an anonymous reviewer for comments that improved this manuscript.

**Open Research**

**Data Availability Statement**

All data used in this study can be found in the National Aeronautics and Space Administration's Planetary Data System (PDS). CRISM, THEMIS, and CTX data used in this study can be directly accessed to PDS Geoscience Node's Mars Data Explorer (https://ode.rsl.wustl.edu/mars/). Laboratory spectral data used in this study was acquired by the Reflectance Experiment Laboratory (RELAB) available through the PDS Geosciences Node Spectral Library (https://pds-speclib.rsl.wustl.edu/). DAFA/TT codes are available on GitHub (https://github.com/linhoml/DAFA-TT-codes-). The code for calculating spectral ratios is available on GitHub (https://github.com/Banus/crism_ml). The location information of pixels containing serpentine in the unprojected CRISM images is available on Zenodo (Emran et al. 2024).

# Global Distribution of Serpentine on Mars

**A. Emran**[1*], **J. D. Tarnas**[2], and **K. M. Stack**[1]

[1] NASA Jet Propulsion Laboratory, California Institute of Technology, CA 91109, USA
[2] Blue Origin, Los Angeles, CA, USA

*Corresponding author: A. Emran (al.emran@jpl.nasa.gov)

**Contents of this file**

    Text S1 to S2

    Figures S1 to S3

    Table S1

## Introduction

This supporting information includes further details about the methodology, i.e., identification of Mg-rich serpentine on Mars using NIR data, the DAFA/TT method, the approach taken for investigating the geologic context and associated mineralogy, and the calculation of $H_2$ produced via serpentinization. Examples of CRISM images and the evaluation approach for assigning the confidence values are provided in Fig. S1, a comparison of the global distribution of serpentine on Mars by this study and Amador et al. (2018) is shown in Fig. S2, and CRISM images with serpentine signatures validated using the radiance ratio method in the Nili Fossae area in Fig. S3. A list of all CRISM images with their assigned confidence values for serpentine signatures is presented in Table S1.

# S1. Methodology

## *S1.1. Identification of Mg-rich Serpentine on Mars*

Serpentine has diagnostic absorption bands in the 1-2.6 µm near-infrared (NIR) range (e.g., King and Clark, 1989). Laboratory spectra of Mg-rich serpentine show a narrow absorption feature at ~1.4-µm, centered on 1.38-µm due to an overtone OH stretching band. Two other characteristic absorption features for Mg-rich serpentine include a narrow asymmetric absorption band at ~2.32-µm due to the Mg-OH combination band and a v-shaped absorption band at ~ 2.52-µm (Clark, 1999). However, the most diagnostic absorption feature for the confident identification of serpentine is a weak, shallow absorption at 2.12-µm due to Mg-OH absorption (Bishop et al. 2008). The distinction of serpentine from other minerals such as carbonate and chlorites on Mars requires the presence of all four absorption bands in the spectra (e.g., Ehlmann et al. 2009, 2010).

The reported the detection using data from the Compact Reconnaissance Imaging Spectrometer for Mars (CRISM) suggests that serpentine is sparse compared to other hydrated minerals, such as phyllosilicates or chlorites (e.g., Ehlmann et al. 2010; Amador et al. 2018). Ehlmann et al. (2010) reports the presence of serpentine in a few regions such as Claritas Rise, Nili Fossae, and near the Isidis basin, and some impact craters in southern highlands. A global scale (70ºN - 70ºS) investigation using the factor analysis/target transformation (FA/TT) method (Malinowski, 1991) showed a total of 49 CRISM images containing serpentine signatures (Amador et al. 2018). However, only six images (FRT0000634B, FRT0000A5AA, FRT0000ABCB, FRT00009971, FRT00003584, and HRL0000B8C2) have been validated for serpentine signatures using radiance ratio technique; five from Amador et al. (2018) and Ehlmann et al. (2010) and one additional image. Leask et al. (2018) validated four of these images, reporting robust spectral signatures of serpentine near Claritas Rise and Nili Fossae, with weaker signatures observed near McLaughlin and Chia craters. The remaining two validated images were confirmed via the radiance ratio method by Lin et al. (2021) and Tarnas et al. (2021).

## *S1.2. DAFA/TT method*

The high rate of false positive serpentine detections using the FA/TT method in visible-to-near-infrared (VNIR) wavelengths (Tarnas et al. 2021) is a limitation of this technique as applied to CRISM data. Moreover, the FA/TT method identifies individual CRISM images that may contain a serpentine deposit but does not localize which pixels contain the serpentine signature (Amador et al. 2018). This has made it difficult to validate serpentine detections from the FA/TT method using the radiance ratio (Leask et al. 2018). An alternate technique, termed dynamic aperture factor analysis and target transformation (DAFA/TT), was developed as an improvement upon the FA/TT method, allowing for the localization of pixels in CRISM image where the serpentine signature originates (Lin et al. 2021). As the DAFA/TT method preserves spatial information about the pixels with detected absorptions, the radiance ratio technique (Leask et al. 2018) can confidently be applied to validate serpentine in CRISM images—distinguishing between true serpentine signatures and artifacts. Consequently, an application of the DAFA/TT method to high-resolution images globally enables the generation of a validated map of Mg-rich serpentine on Mars.

We applied the DAFA/TT method to CRISM images leveraging the high-performance computing (HPC) supercomputer facility at the National Aeronautics and Space Administration (NASA)'s Jet Propulsion Laboratory (JPL). For details of the DAFA/TT method, readers are referred to Lin et al. (2021); however, we summarize the method here. DAFA/TT first derives key signal-eigenvectors or factors from calibrated CRISM data through a method known as factor analysis (Malinowski et al. 1991). To extract these factors, we employ the *Hysime* algorithm (Bioucas-Dias and Nascimento, 2008) with 15 eigenvectors. This particular algorithm and eigenvector number combination were selected based on their proven ability to yield satisfactory results for CRISM images (Tarnas et al. 2021). The 15 signal-eigenvectors are then used for target transformation and compared against a suite of 54 laboratory serpentine spectra sourced from the RELAB database (Milliken et al. 2021)[1] to account for spectral variability in laboratory spectra of serpentine. During target transformation, the signal-eigenvectors and library spectra are modeled using an unconstrained linear least square fit (Keshava and Mustard, 2002), with the root-mean-square error (RMSE) recorded at the pixel level. In this instance, a normalized RMSE threshold value of $1.5 \times 10^{-4}$ was used for positive suggested detections of serpentine. This RMSE value was empirically established as an approximate threshold for detecting serpentine and carbonate in CRISM images using the DAFA/TT method (Lin et al. 2021).

---

[1] https://sites.brown.edu/relab/relab-spectral-database/

The DAFA/TT detections are associated with spatial coordinates in the CRISM image allowing use to extract I/F and radiance data from the candidate pixels. Extracting I/F and radiance spectra from detections involved selecting clusters with an RMSE lower than the specified threshold for at least 20 library serpentine spectra. We generated I/F and radiance ratio images and, subsequently, extracted the corresponding ratio spectra from pixel clusters identified by DAFA/TT. Before ratioing, abrupt spectral spikes in the columns of the calibrated CRISM images were removed following the method of Plebani et al. (2022). In the case of radiance image ratioing, we used either a column median or the same pixels as the denominator in I/F ratioing as the denominator spectra. Visual inspection of I/F ratio spectra focused on characteristics of the serpentine absorption bands at ~ 1.38, 2.12, 2.32, and 2.52-µm. The ratio spectra extracted from DAFA/TT suggested pixels exhibit varying degrees of closeness (similarities) to laboratory serpentine spectra. Consequently, we classified CRISM images with serpentine signatures into confidence values from 1 (low/weak) to 5 (high/best) based on position, strength, and shape of these absorption bands. Accordingly, a confidence value 5 represents a clear match/similarity of extracted CRISM spectra from the candidate pixel cluster to the laboratory serpentine spectra such that the position, strength, and shape of the characteristic absorption bands are stronger and robust. On the hand, the lower the confidence values the weaker the position, strength, and shape of the absorption in the extracted spectra from CRISM data. In the later section of this document (S2), we show the examples of images and the evaluation approach for assigning the confidence values.

*S1.3. Geologic Context and Associated Mineralogy*

The geologic setting of each serpentine detection was visually investigated at a 1:100,000 scale using a THEMIS day IR (Edwards et al. 2011) and CTX global mosaics (Dickson et al. 2023). Investigating the geologic context of the serpentine, we broadly categorized the locational setting into three groups: craters, plains, and valleys. We choose to categorize geologic context into these broad locational settings because the association of serpentine with specific geologic features (e.g., impact craters) constrains the specific mechanisms that could have provided heat and water for serpentinization on Mars. For instance, if serpentine is commonly present in central peaks and/or ejecta of impact craters then the likely formation scenario is that it has formed in the subsurface and was excavated by impact or formed in an impact-generated hydrothermal system. In this instance, it is less likely that the mineral has formed via hydrothermal alteration post-dating impact. On the other hand, if serpentine is only found in mélange terrain the mineral is likely formed via hydrothermal alteration with an uncharacterized heat/water source.

We investigated other minerals contained within the same CRISM image of validated serpentine detections using the standard summary and browse products (Pelkey et al. 2007; Viviano et al. 2014) and a machine learning approach (Plebani et al. 2022). The CRISM summary products of D2300 (Fe/Mg phyllosilicate), BD2500 (carbonate), and OLINDEX3 (olivine) were investigated in this instance. The browse products of MAF (mafic mineralogy) and FM2 (Fe mineral) were used to investigate olivine, while CAR (carbonates) and CR2 (carbonates v2) were used for identifying carbonates including Mg- and Fe/Ca carbonates. For Fe/Mg phyllosilicates, we refer to PFM (phyllosilicates with Fe/Mg) and PHY (phyllosilicates) browse products. I/F spectra were inspected to confirm the presence of other minerals identified in these approaches.

*S1.4. Calculation of $H_2$ Produced via Serpentinization*

We calculate the amount of $H_2$ produced from the surface area covered by validated serpentine deposits (1.46 km$^2$) at different subsurface depths up to ~10 km. We consider this max. depth (~10 km) of serpentinization since the water ceases to flow at that depth (e.g., Oze and Sharma, 2005). Moreover, a hydrothermal fluid-rich region has been estimated to exist >5km deep in the subsurface, with hydrated minerals detected that exhumed from depths exceeding 6km (e.g., Michalski et al. 2013). The rates of $H_2$ production from olivine in laboratory settings show divergent results e.g., ~1 nmol at 70°C (Neubeck et al. 2014) to 30–120 nmol at 100°C (Mayhew et al. 2013) $H_2$ per g of olivine. Furthermore, the composition of olivine (fayalite vs forsterite) controls the production rate of $H_2$ (Klein et al. 2013). The olivine composition on Mars ranges between $Fa_{70}Fo_{30}$ and $Fa_{30}Fo_{70}$ (Hoefen et al. 2003). Since the formation mechanisms, degree of serpentinization, and the temperatures involved with serpentinization are not well constrained on Mars, it is impossible to estimate the exact amount of molecular $H_2$ proceed using the rates (nmol $H_2$/g olivine) given the uncertainty of the variables involved. Moreover, most of these laboratory experiments on $H_2$ production are focused on terrestrial mantle peridotite. In contrast, the basaltic composition of the Martian crust has ~ 2x more iron-content compared to the terrestrial peridotite (Wade et al. 2017). Recently, Tutolo and Tosca (2023) experimentally measured the rate of $H_2$ production per kg rock using Fe-rich olivine from Duluth Complex— a close compositional analog to the Martian crust. Depending on degrees of serpentinization, the rate of $H_2$ production from Fe-rich olivine ranges from 0.01 mol to 0.95 mol per kg rock— a production of ~5x more $H_2$ than the terrestrial peridotite (Tutolo and Tosca, 2023). Being more relevant to Martian crust, we used their minimum and maximum $H_2$ production rates to calculate the ranges of the $H_2$ produced from the detected serpentine deposits. Though Seismic data from the InSight mission indicates a crustal density of 2,850 - 3,100 kg m$^{-3}$ (Wieczorek et al. 2022), we considered the density of "fresh" Martian peridotite to be ~3500 kg m$^{-3}$ (Khan et al. 2021) throughout depth under consideration (~4-10 km). Accordingly, we calculated total $H_2$ production as a function of serpentinization at depth up to the max. of 10 km.

$$H_2 (mol) = A * d * \rho * r \qquad (1)$$

where, *A* is the surface area of serpentine deposits = 1.46 km$^2$, *d* depth at subsurface up to ~10 km, $\rho$ crustal density = 3500 kg m$^{-3}$, and *r* the $H_2$ production rates.

**S2. Examples of Detections**

We show the examples of CRISM images and the evaluation approach for assigning the confidence values of 5 (upper left), 4 (upper right), 3 (lower left), and 2 (lower right) in Fig. S1. For each panel in Fig. S1, the left subplot illustrates the location of the detected serpentine (marked in red spots) overlaid on the CRISM image. The right subplot displays the I/F (blue plot) and radiance (gold plot) ratio spectra extracted from the identified pixels. To facilitate visual comparison of the extracted spectral signature, library spectra of serpentine and other minerals, such as Mg-carbonate and Mg-rich chlorite, with overlapping characteristic absorption bands were used (Fig. S1). These comparisons contribute to the confident assignment of CRISM images to specific values.

*S2.1. McLaughlin Crater (Confidence bin 5)*

The floor of the McLaughlin crater stands out as the location with the highest number of images exhibiting the highest confidence for serpentine detection. Among the seven CRISM images with a confidence value of 5, three of them (FRT0000A5AA, FRT0000A27C, and FRS0002C9DC) are located at the crater. Serpentine spectra from McLaughlin crater (FRT0000A5AA; Fig. S1a) exhibit a strong and broad absorption feature at 2.12-µm in both I/F and radiance ratio spectra, computed as a simple column median ratio. Moreover, both ratio spectra retain the strong absorptions at ~2.32 and 2.52-µm, and an absorption feature at 1.39-µm. The presence of these characteristic absorption bands aligns with features observed in library serpentine spectra. Therefore, the CRISM image at the crater merits a confidence value of 5 for the serpentine signature. While Leask et al. (2018) report a marginal 2.12-µm absorption band in spectra extracted from the same image, we found a robust 2.12-µm absorption feature in both I/F and radiance ratio spectra. The potential reason for this discrepancy in spectral characteristics between Leask et al. (2018) and our study lies in the specific location of the detected pixels in the CRISM image and the resulting ratio spectra.

*S2.2. Nili Fossae (Confidence bin 4)*

The CRISM footprint FRT00003584 at Nili Fossae was previously validated for a serpentine signature by Tarnas et al. (2021). In this study, we validated serpentine spectra from the same image (Fig. S1b). The ratio spectra, calculated through a simple column median ratio, for the detected pixels exhibit a narrow absorption feature at 2.12-µm compared to library spectra (refer to the ratio spectra in Fig S1b). The absorption features at 2.32 and ~1.4-µm align in the right positions, and the spectral shape from 2.3 to 2.5-µm remains consistent in the ratio spectra. Thus, the image was assigned a confidence value of 4 for the serpentine signature.

*S2.3. Chia crater (Confidence bin 3)*

The ratio spectra from the ejecta of a small crater within the Chia crater (HRL000095C7; Fig. S1c) reveals a narrow absorption at 2.12-µm, and the ~1.4-µm band is appropriately positioned. The absorption band at 2.32-µm shows a slight offset from the library spectrum, and the shape from 2.3 to 2.5-µm is slightly different from the library spectrum used in this instance – although there are other serpentine library spectra with variable spectral shapes at these wavelengths. Thus, serpentine detection at the Chia crater was assigned a confidence value of 3. Note that Leask et al. (2018) reported a marginal 2.12-µm absorption in radiance spectra for the same image, while this study indicates a considerable presence of a 2.12-µm feature in both I/F and radiance ratio spectra. This discrepancy arises from the difference in detected pixel locations in the image and the approach to obtaining ratio spectra. While Leak et al. (2018) used a simple column median as the denominator for rationing, we used the nearby "bland pixels" as the denominator (Plebani et al. 2022) for rationing.

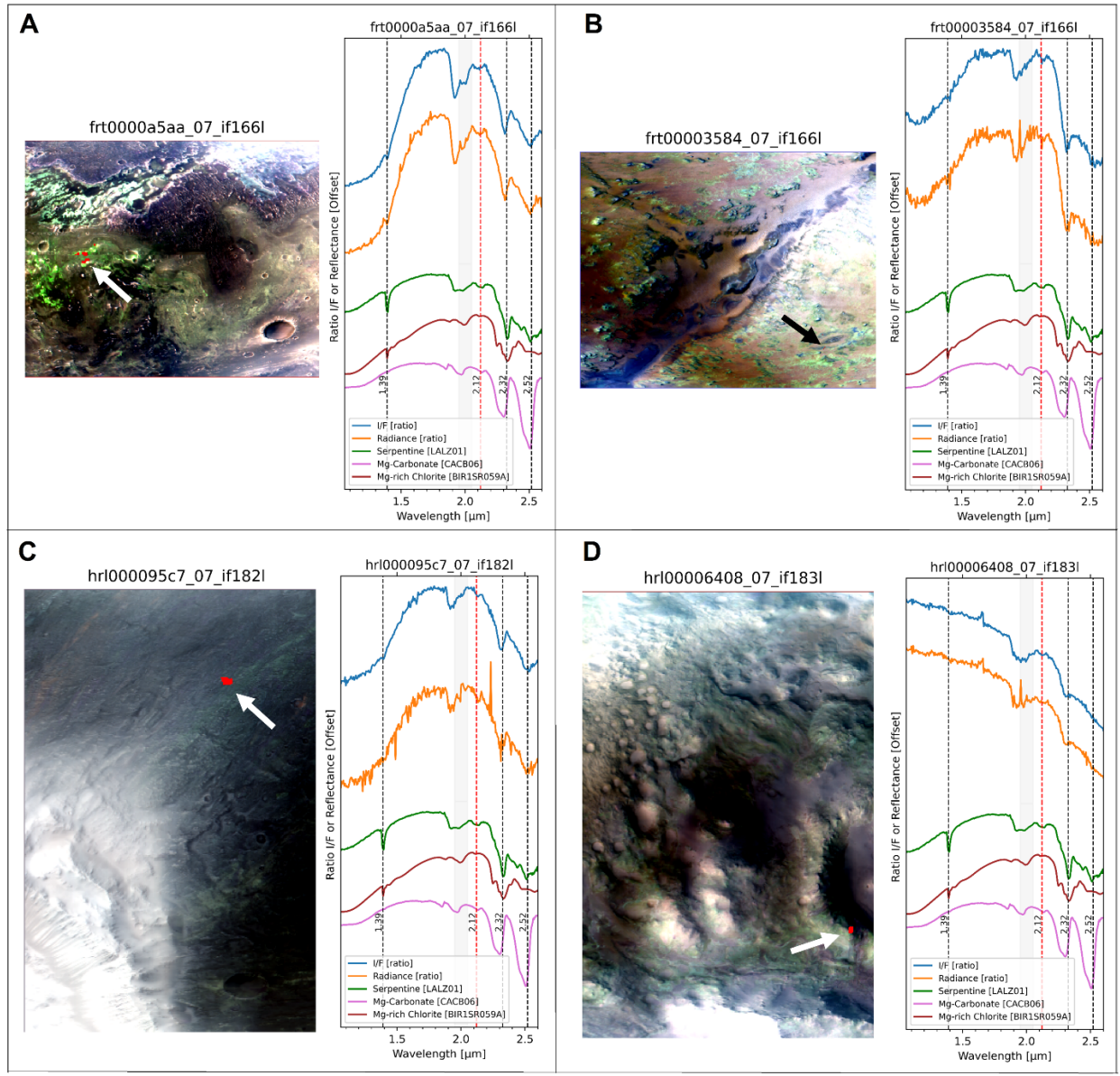

**Fig. S1:** Examples of CRISM images with serpentine confidence values of 5 (a), 4 (b), 3 (c), and 2 (d). In each panel, the left subplot displays the location of the detected serpentine (red spots) on a CRISM image, with arrows (white/black) pointing to the corresponding pixels. The right subplots in each panel depict the I/F (red plot), and radiance (gold plot) ratio spectra extracted from the detected pixels showing presence of 2.12-µm absorption band. Laboratory spectra from the RELAB database for serpentine (green plot), Mg-carbonate (pink plot), and Mg-rich chlorite (marron plot) are used for comparison. The digits on the parentheses following the name of library spectra in the legend indicate the specimen ID in the database. Refer to the text in the appropriate section of this document for the detailed interpretation of each subplot.

*S2.4. Chaos Terrains near Nili Fossae (Confidence bin 2)*

Serpentine was detected in the chaos terrain at the dichotomy boundary near Nili Fossae within the CRISM footprint HRL00006408 (Fig. S1d). The ratio spectra, calculated through a simple column median ratio, extracted for the detected pixels exhibit a very weak but broad absorption feature at 2.12-µm, considering the noise level in the spectra. The absorption feature at 1.4-µm is weak, the 2.3-µm band has right position but has slightly offset shape, the shape from ~2.3 to 2.5-µm is not entirely consistent with laboratory spectra. Consequently, based on noise level in the spectra and strengths of diagnostic absorption bands, the image was assigned a confidence level of 2 for the serpentine signature.

*S2.5. Concentration at Nili Fossae and other regions*

Near the original proposed landing site ellipse of the Mars Science Laboratory (MSL) in the Nili Fossae trough floor (Fig. S3), four images (FRT00008FC1, FRT00004F75, FRS000355D1, and FRS00029EA8) show a medium confidence value of 3. Two other images (FRT00009971 and FRT000088D0), located a little distant from the landing ellipse, show a confidence value of 4. The central peak (remnant) or crater fill at Leighton crater in the Tyrrhenian Terra (southwest of Isidis Planitia) exhibits two images with confidence values of 3 (FRT00009265) and 2 (FRT0000A546). The cratered terrains north of Hellas Planitia show serpentine signatures in two images (FRT0000C58A and FRT00008144), both with a confidence value of 3.

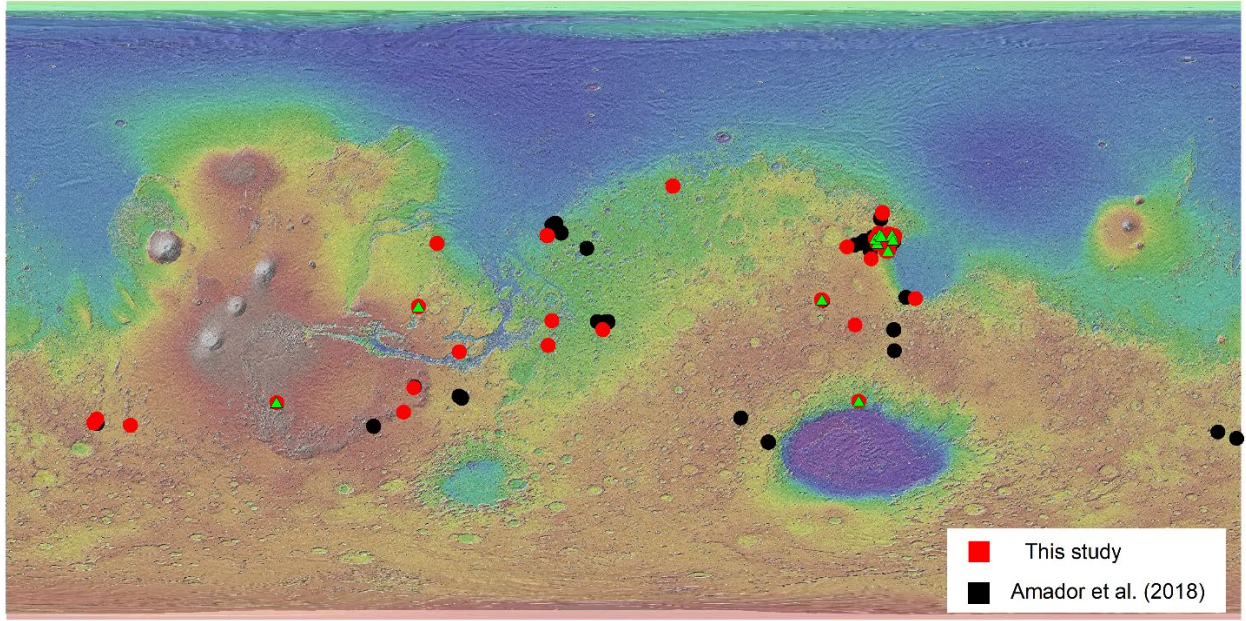

**Fig. S2**: Comparison of the global distribution of serpentine on Mars by this study and Amador et al. (2018) overlain MOLA shaded elevation data. The black and red circles represent detection by Amador et al. (2018) and this study, respectively. The green triangles represent location of CRISM images from Amador et al. (2018) that were validated for characteristic serpentine absorptions using the radiance ratio technique in this study.

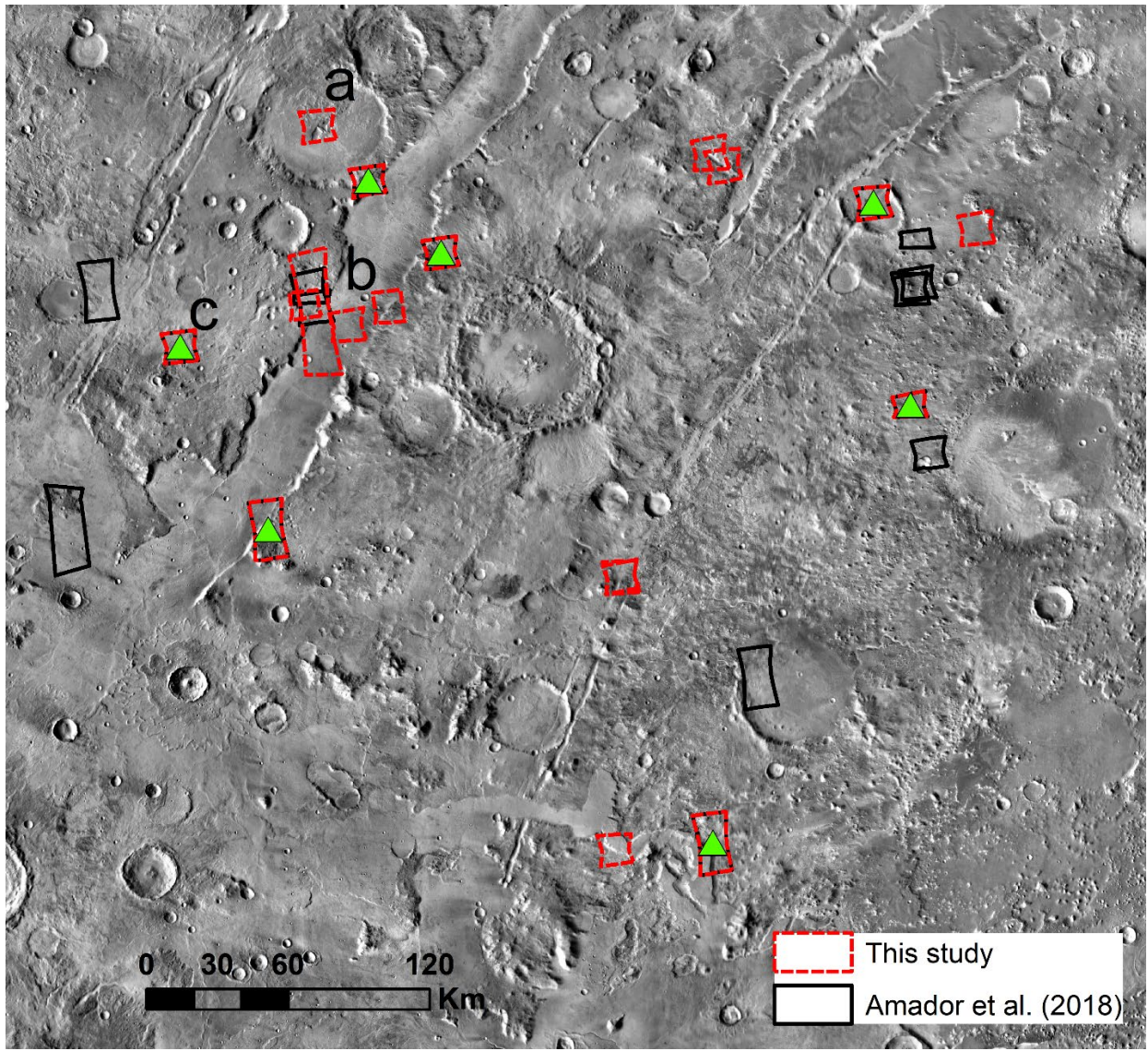

**Fig. S3**: CRISM images with serpentine signature validated using radiance ratio method overlain THEMIS IR-daytime mosaic at Nili Fossae area. A total of 18 CRISM images were verified in the region in this study (dashed red outlines), while Amador et al. (2018) has also identified 18 candidate images consistent with the serpentine (solid black outlines). CRISM footprints centered with green triangles represent images from Amador et al. (2018) validated for characteristic serpentine absorptions using the radiance ratio technique. Refer to the text in the appropriate section of the article for the details of confidence values. The letters a, b, and c present serpentine detection in CRISM footprint that occurs in crater, valley, and plains, respectively.

## S3. List of all detected images

**Table S1:** A list of all CRISM images with their assigned confidence value for serpentine signature, the method of choosing denominator spectra for rationing (simple column median vs nearby "bland" pixels), along with other associated information.

| CRISM_ID | Region | Geologic context | Confidence | Ratio method | Age | Fe-Mg Phyl. | Carbonate | Olivine | Other minerals |
|---|---|---|---|---|---|---|---|---|---|
| HRL0000B8C2_07* | Nili Fossae | Rough plains | 2 | Median | N | x | x | x | Jarosite, Kaolinite, Low-Ca pyroxene |
| HRL000095C7_07* | Xanthe Terra/ Valles Marineris | Inside the Chia crater | 3 | Bland | N | x | x | x | Chlorite, Low-Ca pyroxene |
| HRL000095A2_07* | Nili Fossae | Intercrater plains | 5 | Median | N | x | x | x | Kaolinite, Low-Ca pyroxene |
| HRL0000860C_07 | Terra Sirenum | Crater ejecta/ knobby terrain | 3 | Bland | N | x | x | x | Chloride, Low-Ca pyroxene |
| HRL00006408_07 | Nili Fossae | Crater rim and wall | 2 | Median | HN | x | x | x | Jarosite, Al smectite, Low-Ca pyroxene, Kaolinite |
| FRT0001957D_07 | Lunae Planum | Knobby terrain | 2 | Bland | H | x | x | | Low-Ca pyroxene |
| FRT00018AF5_07 | Margaritifer Terra | Intercrater plains | 2 | Bland | N | x | | x | Low-Ca pyroxene |
| FRT00018154_07 | Thaumasia Planum | Smooth plains | 4 | Bland | N | x | x | x | Low-Ca pyroxene |
| FRT0001406E_07 | Margaritifer Terra | Intercrater plains | 3 | Median | H | x | | | |
| FRT0000CBE5_07 | Nili Fossae | Intercrater plains | 3 | Bland | N | x | x | x | Kaolinite, Al-smectite, Low-Ca pyroxene |
| FRT0000C968_07 | Nili Fossae | Smooth plains | 3 | Median | N | x | x | x | |
| FRT0000C58A_07 | N. Hellas Planitia | Crater rim and wall | 3 | Median | N | x | x | x | Chlorite, Low-Ca pyroxene |
| FRT0000C380_07 | N Arabia Terra | Smooth plains | 2 | Median | AN | x | | x | Hydrated silica, Jarosite |
| FRT0000B438_07 | Nili Fossae | Small channel wall | 3 | Bland | N | x | x | x | Low-Ca pyroxene |
| FRT0000ABCB_07* | Nili Fossae | Crustal mélange | 5 | Median | N | x | x | x | Kaolinite, Low-Ca pyroxene, Hydrated silica |
| FRT0000AB81_07 | Terra Sirenum | Knobby terrain/ polygonal cracks | 2 | Bland | N | x | x | x | Chloride |
| FRT0000A819_07 | Tyrrhena Terra/S. Isidis | Intercrater plains | 3 | Bland | N | x | x | x | Low-Ca pyroxene |
| FRT0000A5AA_07 | Mawrth Vallis | Crater floor at McLaughlin crater | 5 | Median | N | x | x | x | Low-Ca pyroxene |
| FRT0000A546_07* | Tyrrhena Terra/S. Isidis | Central peak (remnant) at Leighton crater | 2 | Bland | AH | x | x | x | Chlorite, Illite, Kaolinite, Hydrated silica, Low-Ca pyroxene |
| FRT0000A2B3_07 | Nili Fossae | Crater central peak | 4 | Bland | H | x | x | x | Hydrated silica, Chlorite, Jarosite, Illite, Al smectite, Low-Ca pyroxene |
| FRT0000A27C_07 | Mawrth Vallis | Crater floor at McLaughlin crater | 5 | Median | N | x | x | x | Low-Ca pyroxene |
| FRT0000A053_07 | Nili Fossae | Crater floor (fill) | 2 | Bland | N | x | x | x | Kaolinite, Jarosite, Hydrated silica, Al smectite, Low-Ca pyroxene |
| FRT00009ACE_07 | SW Meridiani Planum | Intercrater plains | 2 | Median | N | x | | x | |
| FRT00009AAA_07 | Sirenum Tholus | Intercrater plains | 3 | Bland | N | x | | x | Low-Ca pyroxene |
| FRT00009971_07* | Nili Fossae | Western wall of Nili Fossae trough | 4 | Bland | N | x | x | x | Kaolinite, Chlorite, Hydrated silica, Al smectite, Low-Ca pyroxene |
| FRT000097E2_07 | Nili Fossae | (Exposed) plains | 3 | Bland | N | x | x | x | Kaolinite, Hydrated silica, Prehnite, Low-Ca pyroxene, |
| FRT000093BE_07 | Nili Fossae | (Exposed) plains | 4 | Median | N | x | x | x | Kaolinite, Hydrated silica, Al-smectite, Low-Ca pyroxene |
| FRT00009265_07 | Tyrrhena Terra/S. Isidis | Central peak (remnant) at Leighton crater | 3 | Bland | AH | x | x | x | Chlorite, Illite, Low-Ca pyroxene |
| FRT00008FC1_07 | Nili Fossae | Crater ejecta | 3 | Median | AH | x | x | x | Kaolinite, Low-Ca pyroxene |
| FRT000088D0_07* | Nili Fossae | Crater ejecta | 4 | Bland | AH | x | x | x | Kaolinite, Hydrated silica, Illite, Prehnite, Chlorite, Low-Ca pyroxene |
| FRT00008144_07* | N. Hellas Planitia | Crater rim and wall | 3 | Bland | N | x | x | x | Chlorite, Low-Ca pyroxene |
| FRT00008112_07 | Capri Chasma at Valles Marineris | Northern wall of Capri Mensa | 3 | Median | N | x | x | x | Chlorite, Low-Ca pyroxene |
| FRT00006485_07 | Tyrrhena Terra/ S. Syrtis Major | Crater central peak | 4 | Bland | N | x | x | x | Chlorite, Hydrated silicate, Illite |
| FRT0000634B_07* | Claritas Rise | Knobby or mélange terrain | 5 | Median | N | x | | | Illite, Chlorite, Kaolinite, Hydrated silica, Prehnite, Low-Ca pyroxene |
| FRT000050F2_07 | Syrtis Major | Crater wall | 5 | Median | H | x | x | x | Chlorite, Kaolinite, Prehnite, Low-Ca pyroxene |
| FRT00004F75_07 | Nili Fossae | Nili Fossae trough floor | 3 | Median | N | x | x | x | Kaolinite, Low-Ca pyroxene |
| FRT00003E12_07 | Nili Fossae | (Exposed) plains | 3 | Bland | N | x | x | x | Hydrated silica, Low-Ca pyroxene |

| CRISM Image ID | Region | Geologic context | Robustness | Ratio method | Age | | | Associated minerals |
|---|---|---|---|---|---|---|---|---|
| FRT00003584_07* | Nili Fossae | Crater floor | 4 | Median | N | x | x | Hydrated silica, Low-Ca pyroxene |
| FRT000028BA_07* | Nili Fossae | Rough plains | 3 | Bland | HN | x | x | Low-Ca pyroxene |
| FRS0003F22E_01 | Thaumasia Planum | Vally wall/ slumping material | 4 | Bland | N | x | | Kaolinite, Low-Ca pyroxene |
| FRS000355D1_01 | Nili Fossae | Crater ejecta | 3 | Median | AH | x | | Low-Ca pyroxene |
| FRS0002C9DC_01 | Mawrth Vallis | Crater floor at McLaughlin crater | 5 | Bland | N | x | | Low-Ca pyroxene |
| FRS00029EA8_01 | Nili Fossae | Nili Fossae trough floor | 3 | Bland | N | x | | Kaolinite, Low-Ca pyroxene |

***Note:***
- CRISM image ID with asterisk (*) indicates the image detected by Amador et al. (2018) and validated using radiance ratio technique (Leask et al. 2018) in this study.
- The geologic context of the detected serpentine is evaluated at 1: 100,000 scale using CTX and THEMIS daytime mosaic.
- Robustness index: 5 is best (robust) spectra and 1 is weak spectra.
- Ratio method: Selection of denominators for rationing. Median = simple median from the same detector column and bland = nearby "bland pixels" following the method described in the method section.
- Age column codes: A = Amazonian, AH= Amazonian-Hesperian, AN = Amazonian- Noachian, H = Hesperian, HN = Hesperian-Noachian, and N = Noachian.